\documentclass[%
 reprint,
 amsmath,amssymb,
 aps,
superscriptaddress, 
]{revtex4-2}

\usepackage{appendix}
\usepackage{graphicx}% Include figure files
\usepackage{dcolumn}% Align table columns on decimal point
\usepackage{bm}% bold math
\begin{document}

% \preprint{APS/123-QED}

\title{Fundamental bounds on the precision of classical phase microscopes}

\author{Dorian Bouchet}
\thanks{These authors contributed equally.}
  \affiliation{Debye Institute for Nanomaterials Science, Utrecht 
University, P.O. Box 80000, 3508 TA Utrecht, the Netherlands}%Lines 
\author{Jonathan Dong}%
\thanks{These authors contributed equally.}
  \affiliation{Laboratoire Kastler Brossel, Ecole Normale 
Supérieure, Université PSL, CNRS, Sorbonne Université, Collège de 
France, F-75005 Paris, France}
\author{Dante Maestre}%
  \affiliation{University of Vienna, Faculty of Physics, VCQ, 
A-1090 Vienna, Austria}
  \affiliation{University of Vienna, Max Perutz Laboratories, 
Department of Structural and Computational Biology, A-1030 Vienna, 
Austria}

\author{Thomas Juffmann}
\thanks{Corresponding author: thomas.juffmann@univie.ac.at}
  \affiliation{University of Vienna, Faculty of Physics, VCQ, 
A-1090 Vienna, Austria}
  \affiliation{University of Vienna, Max Perutz Laboratories, 
Department of Structural and Computational Biology, A-1030 Vienna, 
Austria}

% custom shortcut
\newcommand{\bra}{\langle}
\newcommand{\ket}{\rangle}
\newcommand{\Ee}{\operatorname{E}}
\newcommand{\Ve}{\operatorname{Var}}
\newcommand{\Med}{\operatorname{Med}}
\newcommand{\de}{\operatorname{d}}
\newcommand{\re}{\operatorname{Re}}
\newcommand{\im}{\operatorname{Im}}
\newcommand{\tr}{\operatorname{Tr}}
\newcommand{\refr}[1]{Ref.~\cite{#1}}
\newcommand{\refrs}[2]{Ref.~\cite{#1} and~\cite{#2}}
\newcommand{\eg}{e.g.\@\xspace}
\newcommand{\cf}{cf.\@\xspace}
\newcommand{\eq}[1]{Eq.~\eqref{#1}}
\newcommand{\eqsb}[1]{Eqs.~\eqref{#1}}
\newcommand{\eqs}[2]{Eqs.~\eqref{#1} and~\eqref{#2}}
\newcommand{\eqss}[3]{Eqs.~\eqref{#1}, \eqref{#2} and~\eqref{#3}}
\newcommand{\eqsss}[4]{Eqs.~\eqref{#1}, \eqref{#2}, \eqref{#3} and~\eqref{#4}}
\newcommand{\Eq}[1]{Equation~\eqref{#1}}
\newcommand{\Eqs}[2]{Equations~\eqref{#1} and~\eqref{#2}}
\newcommand{\fig}[1]{Fig.~\ref*{#1}}
\newcommand{\figs}[2]{Figs.~\ref*{#1} and~\ref*{#2}}
\newcommand{\Fig}[1]{Figure~\ref*{#1}}
\newcommand{\tab}[1]{Table~\ref*{#1}}
\newcommand{\textgreek}[1]{\begingroup\fontencoding{LGR}\selectfont#1\endgroup}

\renewcommand{\eqref}[1]{(\ref{#1})}

\date{\today}% It is always \today, today,

\begin{abstract}
A wide variety of imaging systems have been designed to measure phase variations, with applications from physics to biology and medicine. In this work, we theoretically compare the precision of phase estimations achievable with classical phase microscopy techniques, operated at the shot-noise limit. We show how the Cram\'{e}r-Rao bound is calculated for any linear optical system, including phase-contrast microscopy, phase-shifting holography, spatial light interference microscopy, and local optimization of wavefronts for phase imaging. Through these examples, we demonstrate how this general framework can be applied for the design and optimization of classical phase microscopes. Our results show that wavefront shaping is required to design phase microscopes with optimal phase precision. 
\end{abstract}

\maketitle

\section{Introduction}
Phase microscopy enables precise measurements of optical path length differences~\cite{Zernike1942, Gabor1948APrinciple,Popescu2011,Lee2013QuantitativeApplications}. Any phase image is constructed from a finite number of detected photons. This number can be limited by technological constraints, such as the source power or the dynamic range of the detector. The maximum photon flux can also be dictated by specimen damage, e.g. in dispersive imaging of ultra-cold atoms where inelastic scattering and elastic recoil lead to atom loss and heating~\cite{Andrews1996DirectCondensate, Windpassinger2008NondestructiveLimit}, respectively. Another example is the imaging of biological structures with ultraviolet (UV) radiation~\cite{Ojaghi2018UltravioletMicroscopy, Cheung2013IntracellularMapping}, where inelastic channels lead to sample damage. In such applications, it is therefore crucial to maximize the information retrieved per photon. 

Nowadays, many different variations of phase microscopes exist~\cite{Popescu2011}, all of which convert optical path length differences into detectable intensity variations. The Cram\'{e}r-Rao bound (CRB) allows quantifying the precision that can be achieved with any given phase-microscopy technique, as it imposes a lower limit to the precision that can be achieved in the estimation of parameters from noisy data~\cite{Trees2013DetectionI,Barrett2013FoundationsScience}. Such quantitative analysis of the achievable estimation precision is of great importance in many fields, be it in fluorescence microscopy regarding particle localization precision~\cite{Deschout2014,Shechtman2014OptimalImaging,balzarotti_nanometer_2017}, in interferometric scattering microscopy regarding the possibility to detect and characterize single proteins without labelling them~\cite{Piliarik2014, bouchet2020influence}, or in electron microscopy regarding particle structure determination~\cite{doerschuk2000ab}. In the context of phase microscopy, several recent studies have focused on the question of how to increase the precision of phase estimations using quantum correlations of the probe light~\cite{Humphreys2013QuantumEstimation, Liu2020QuantumEstimation}. However, quantum enhanced schemes are difficult to realize experimentally, and common phase microscopes do not even reach the optimum CRB achievable with uncorrelated light. 

Here, we expose the experimental conditions necessary to reach the optimal precision on phase estimations based on shot-noise limited measurements. Via a Cram\'{e}r-Rao analysis, we demonstrate that common phase-microscopy techniques fall short of this limit because they are not ideal for the specimens under study. Additionally, we show that wavefront-shaping techniques, which recently found several applications in phase microscopy~\cite{ Maurer2011,Wang2011SpatialSLIM,Toda2020AdaptiveImaging,Bouchet2020MaximumMeasurements,Juffmann2018}, can be used to recover the optimal estimation precision for any given phase sample. The manuscript is organized as follows. In Section~\ref{sec:theoretical}, we derive fundamental bounds based on the Cram\'er-Rao inequality for the precision of any classical linear phase microscope. In Section~\ref{sec:externalref}, we discuss how to reach this bound in the case of an external reference, which requires wavefront shaping for strong phase objects. In Section~\ref{sec:internalref}, we show how this framework also applies to imaging systems with an internal reference.

%More specifically, we discuss interferometric imaging with either external or internal reference waves, comparing schemes in which the phase between the reference wave and the average object wave is either fixed (e.g. phase contrast microscopy or PCM \cite{Zernike1942})), varied (e.g. off-axis holography \cite{Leith1963WavefrontObjects}, spatial light interference microscopy (SLIM \cite{Popescu2004})), or locally adapted to the specimen (e.g. local optimization of wavefronts for phase imaging (LowPhi, \cite{Juffmann2018})). We address the complementarity of phase and absorption measurements, imaging artefacts, as well as specimens which cannot be considered a weak phase object.

\section{Theoretical framework}\label{sec:theoretical}

\subsection{General model for phase microscopes}

We consider a general model in which coherent light propagates through a phase sample, so that local phase shifts induced by the object are imprinted in the transmitted wavefront. By partitioning our region of interest in $p$ small surfaces of area $S_{\mathrm{a}}$, the average number of photons passing through the $j$-th area within a time $\Delta t $ is then given by $n_j =\Delta t |E^{\mathrm{obj}}_j|^2$, where $E^{\mathrm{obj}}_j=\mathcal{A}_j e^{i \phi_j}$ denotes the complex field associated with this area. Note that this definition of the field differs from the classical definition of the complex electric field (in SI units) by a factor $\sqrt{(n_{\mathrm{obj}}\epsilon_0 c S_{\mathrm{a}})/(2  \hbar  \omega)} $ where $n_{\mathrm{obj}}$ is the object refractive index, $\epsilon_0$ is the vacuum permittivity, $c$ is the speed of light in vacuum, $\omega$ is the photon angular frequency, and $\hbar$ is the reduced Planck constant. 

To estimate these phase values $\phi=(\phi_1, \dots, \phi_p)$, we use a linear optical system characterized by an operator $H \in \mathbb{C}^{d \times p}$ whose elements are noted $h_{kj}=|h_{kj}|e^{i \beta_{kj}}$. With these notations, the average intensity image on the camera is 
\begin{equation}
I^\mathrm{det} =  \left \vert H E^{\rm{obj}} + E^{\rm{ref}} \right \vert^2 \; ,
\label{expected_intensity}
\end{equation} 
where we introduced an optional external reference field $E^{\rm{ref}} \in \mathbb{C}^d$.

Let us consider that the data measured by the camera are described by a $d$-dimensional random variable $X$ characterized by a joint density probability function $p(X;\phi)$. The variance of any unbiased estimator $\hat{\phi}(X)$ of $\phi$ must satisfy the Cram\'{e}r-Rao inequality~\cite{Trees2013DetectionI}, which is expressed by
\begin{equation}
 \operatorname{Var}(\hat{\phi}_j) \geq [ \mathcal{J}^{-1} (\phi) ]_{jj} \; ,
 \label{CRLB_general}
\end{equation}
where $\operatorname{Var}$ is the variance operator and $\mathcal{J}(\phi)$ the Fisher information matrix defined by
\begin{equation}
 [\mathcal{J}(\phi)]_{ij} = \operatorname{E} \left[\left( \frac{\partial \ln p(X;\phi)}{\partial \phi_i}\right) \left( \frac{\partial \ln p(X;\phi)}{\partial \phi_j}\right) \right] \; ,
 \label{fisher_definition}
\end{equation}
with $\operatorname{E}$ the expectation operator acting over noise fluctuations. Maximum-likelihood estimators are known to saturate this inequality in the asymptotic limit~\cite{Trees2013DetectionI}. Consequently, measuring a large number of photons is a sufficient condition for this inequality to be saturable. 

Since the Fisher information matrix is positive semi-definite, we can also write the following inequality:
\begin{equation}
 [ \mathcal{J}^{-1} (\phi) ]_{jj} \geq \left[ \mathcal{J} (\phi) \right]_{jj}^{-1} .
 \label{eq: several CRB inequalities}
\end{equation}
This inequality, which results from properties of the Schur complement, highlights the central role played by the diagonal elements of the Fisher information matrix upon the precision of estimations. It is in fact saturated when the Fisher information matrix is diagonal, which corresponds to the case in which the estimation of $\phi_j$ is not influenced by an imperfect knowledge of all other parameters.  Alternatively, this also corresponds to the case in which one seeks to estimate a given parameter $\phi_j$ assuming that the value of all other parameters is known. 

\subsection{Fisher information in the shot-noise limit}

We now assume that shot noise, i.e. the detection statistics due to the quantized nature of light, is the main source of noise, and that we can neglect mechanical vibrations, readout noise, dark currents, and other sources of noise. Considering an integration time $\Delta t$ and assuming that the values measured by all camera pixels are statistically independent, the probability density function $p(X;\theta)$ simplifies to
\begin{equation}
 p(X;\theta) = \prod_{k=1}^d \frac{e^{- \Delta t I_k^{\rm{det}}} (\Delta t I_k^{\rm{det}})^{X_k}}{X_k!} \; . 
 \label{poisson_statistics}
\end{equation}
Inserting Eq.~\eqref{poisson_statistics} into Eq.~\eqref{fisher_definition} yields
\begin{equation}
 \label{general FI single parameter}
 [\mathcal{J}(\phi)]_{ij} 
 = \Delta t
 \sum_{k=1}^d
 \frac{1}{I_k^{\rm{det}}}
 \left( \frac{\partial I_k^{\rm{det}}}{\partial \phi_i} \right)
 \left( \frac{\partial I_k^{\rm{det}}}{\partial \phi_j} \right) \; .
\end{equation}
Using the equality $I^{\rm{det}} = |E^{\rm{det}}|^2$ to express the derivative of the intensity $I_k^\mathrm{det}$ with respect to $\phi_j$, we obtain
\begin{align}
 \label{derivative of intensity 2}
 \frac{\partial I_k^{\rm{det}}}{\partial \phi_j} 
 = - 2\, \text{Im}\left[ (E^{\rm{det}}_k)^* h_{kj} E^{\rm{obj}}_j \right] \; .
\end{align}
% The elements of the Fisher information matrix can thus be explicitly expressed as follows:
% % \begin{equation}
% %  \label{explicit fisher information matrix}
% %  [\mathcal{J}(\phi)]_{ij}
% %  = 4 \Delta t
% %  \sum_{k=1}^d \frac{ \text{Im} \left[ (E^{\rm{det}}_k)^* h_{ki} E^{\rm{obj}}_i \right]
% %  \text{Im}\left[ (E^{\rm{det}}_k)^* h_{kj} E^{\rm{obj}}_j \right]}{|E^{\rm{det}}_k|^2} \; .
% % \end{equation}
% \begin{equation}
%  \label{explicit fisher information matrix}
%  \begin{split}
%  [\mathcal{J}(\phi)]_{ij}
%  = 4 \Delta t
%  \sum_{k=1}^d &\frac{1}{|E^{\rm{det}}_k|^2} \text{Im} \left[ (E^{\rm{det}}_k)^* h_{ki} E^{\rm{obj}}_i \right] \\
%  &\times \text{Im}\left[ (E^{\rm{det}}_k)^* h_{kj} E^{\rm{obj}}_j \right] \; .
%  \end{split}
% \end{equation}
Writing $E^{\rm{det}}_k = |E^{\rm{det}}_k| e^{i \alpha_k} $, the diagonal elements of the Fisher Information matrix are expressed by
\begin{equation}
 [\mathcal{J}(\phi)]_{jj}= 4 n_j \sum_{k=1}^d |h_{kj}|^2 \sin^2(\phi_j + \beta_{kj} - \alpha_k) \; . 
 \label{diag_elements}
\end{equation}

Equation \eqref{diag_elements} is maximized if two conditions are satisfied. First, we have $ \sin^2(\phi_j + \beta_{kj} - \alpha_k) \leq 1$, with the equality holding if the following phase-matching condition is fulfilled:
\begin{equation}
 \label{eq: phase condition with external reference}
 \phi_j + \beta_{kj} - \alpha_k = \pi / 2 + m\pi \; ,
\end{equation}
where $m$ can be any integer. The condition must hold for all $k, j$ for which $n_j h_{kj} \neq 0$.
Second, we can remark that energy conservation imposes a condition on $H$. Indeed, the total intensity of the object field is $\|E^{\rm{obj}}\|^2$, and becomes after propagation through the optical system $\|H E^{\rm{obj}}\|^2 = E^{\rm{obj}\dag} H^\dag H E^{\rm{obj}}$. 
In general, since the passive linear optical system can only decrease the transmitted energy, the inequality $\|H E^{\rm{obj}}\|^2 \leq \|E^{\rm{obj}}\|^2$ is always satisfied. % Moreover, when the transformation applied by the optical system is unitary, $H^\dag H$ is the identity matrix of size $p$ and energy is conserved. 
Energy conservation thus imposes that $\sum_k |h_{kj}|^2 \leq 1$, with the equality holding if energy is conserved and $H$ is a unitary matrix. Hence, %considering that $ \sin^2(\phi_j + \beta_{kj} - \alpha_k) \leq 1$ and $\sum_k |h_{kj}|^2 \leq 1$, 
we obtain the following inequality:
\begin{equation}
 [\mathcal{J}(\phi)]_{jj}\leq 4 n_j \; ,
 \label{diag_elements_bound}
\end{equation}
which is saturated if $H$ is unitary and if the phase-matching condition expressed by Eq.~\eqref{eq: phase condition with external reference} is satisfied.

To summarize the results from this section, we can use Eqs.~\eqref{CRLB_general}, \eqref{eq: several CRB inequalities} and \eqref{diag_elements_bound} to write the following chain of inequalities:
\begin{equation}
 \operatorname{Var}(\hat{\phi}_j) \geq [ \mathcal{J}^{-1} (\phi) ]_{jj} \geq [ \mathcal{J} (\phi) ]_{jj}^{-1} \geq \frac{1}{4 n_j} \; .
 \label{eq: inequ}
\end{equation}

In order to design a phase microscope that reaches this fundamental bound, one needs to saturate all the inequalities in Eq.~(\ref{eq: inequ}), leading to several conditions. The first inequality is saturated asymptotically with maximum-likelihood estimators~\cite{Trees2013DetectionI}, which implies that this inequality can be saturated when $\Delta t I_k^{\rm{det}}\gg 1$ over the whole field of view (FOV). To saturate the second inequality, the Fisher information matrix needs to be diagonal, such that no correlation exists in the estimated values of different parameters. For the last inequality to be saturated, two additional conditions must be satisfied. First, the optical system has to be lossless and thus represented by a unitary matrix $H$. Second, for each camera pixel $k$ that satisfies $|h_{kj}|^2 \neq 0$, the following phase-matching condition needs to be fulfilled:
\begin{equation}
\phi_j + \beta_{kj} - \alpha_k =  \pi/2 + m\pi \; ,
\label{eq:phase_condition}
\end{equation}
 where $\alpha_k = \arg(E_k^{\mathrm{det}})$ denotes the phase of the field in the camera plane.

Using the same assumptions, an analogue of Eq.~\eqref{eq: inequ} can also be derived for the estimation of other parameters. A common application is for example when one needs to estimate the number of photons absorbed by the object. Absorptive samples can indeed modulate the amplitude of the incident field, and one may want to estimate the set of parameters $n = (n_1, \ldots, n_p)^{\mathsf{T}}$ instead of $\phi$. In such case, we obtain the following chain of inequalities (see Supplementary Section~S1): 
\begin{equation}
\operatorname{Var} (\hat{n}_j ) \geq [ \mathcal{J}^{-1} (n) ]_{jj}  \geq [ \mathcal{J} (n) ]_{jj}^{-1}  \geq n_j \; .
\end{equation} 
The conditions required to saturate these inequalities are essentially the same as the ones introduced for phase estimations, except that the phase-matching condition now reads 
\begin{equation}
    \phi_j + \beta_{kj} - \alpha_k = m\pi \; .
\end{equation}
There is an apparent trade-off between phase and absorption estimations: no absorption information is available when phase information is optimally captured by the measurements, and vice versa.

\begin{figure}
    \centering
    \includegraphics[width=\columnwidth]{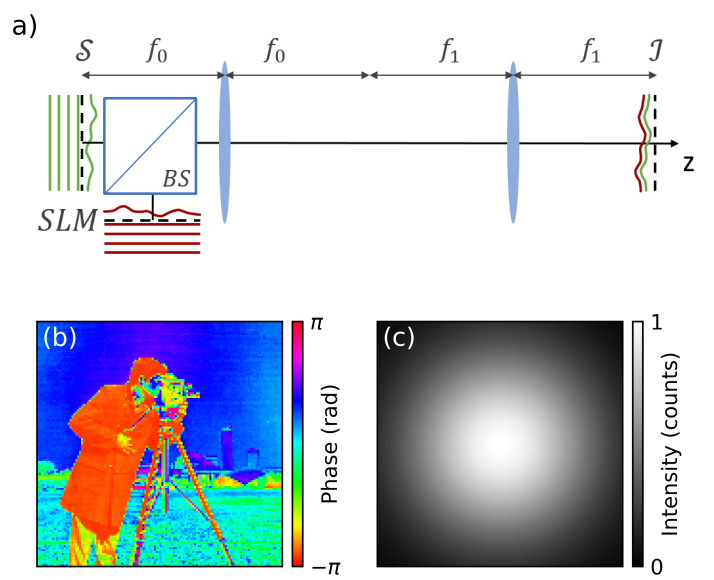}
    \caption[setup]{(a)~Schematics of a phase microscope operating with external reference wave. The sample $\mathcal{S}$ is imaged onto the image plane $\mathcal{I}$ using a 4f setup. The object wave (green) is interfered with an external reference (red), which is in-coupled using a beam splitter (BS). The reference wave can be shaped using a spatial light modulator (SLM). (b)~Phase of the object wave after interacting with an artificial non-absorbing sample ranging from $-\pi$ to $\pi$. (c)~Intensity of the object wave, showing the Gaussian envelope of the incoming beam.} 
    \label{fig:setup}
\end{figure}

\section{Phase microscopes with an external reference}\label{sec:externalref}

We first study the consequences of this phase-matching condition for phase microscopes with an external reference. In a simple configuration, the incident wave passes through the sample and is imaged onto the camera using an ideal 4f system represented by an identity $H$-matrix (Fig.~\ref{fig:setup}a). In the detector plane, the field interferes with a high-intensity external reference beam ($|E^\mathrm{ref}_k|^2/|E^\mathrm{obj}_k|^2=100$ in the center of the FOV), whose phase profile can be shaped using a noiseless spatial light modulator (SLM). The two beams are coupled using a beam-splitter with a transmission close to $1$ to preserve the unitarity of $H$.
With such phase microscope, the phase-matching condition expressed by \eq{eq:phase_condition} reduces to $\phi_k - \alpha_k = \pi / 2 + m\pi$ where $\alpha_k \simeq \arg(E_k^{\mathrm{ref}})$. Thus, if the reference field is a plane wave with a phase that is shifted by $\pi/2$ with respect to the average phase of the object wave (assumed to be 0 for simplicity), the phase-matching condition is fulfilled for a weakly-contrasted object ($\phi_j \ll \pi/2$), yielding an optimal precision for phase estimations. However, this strategy cannot be applied to strongly-contrasted objects, as typically encountered in cell biology. 

\begin{figure}
	\centering
	\includegraphics[width=\columnwidth]{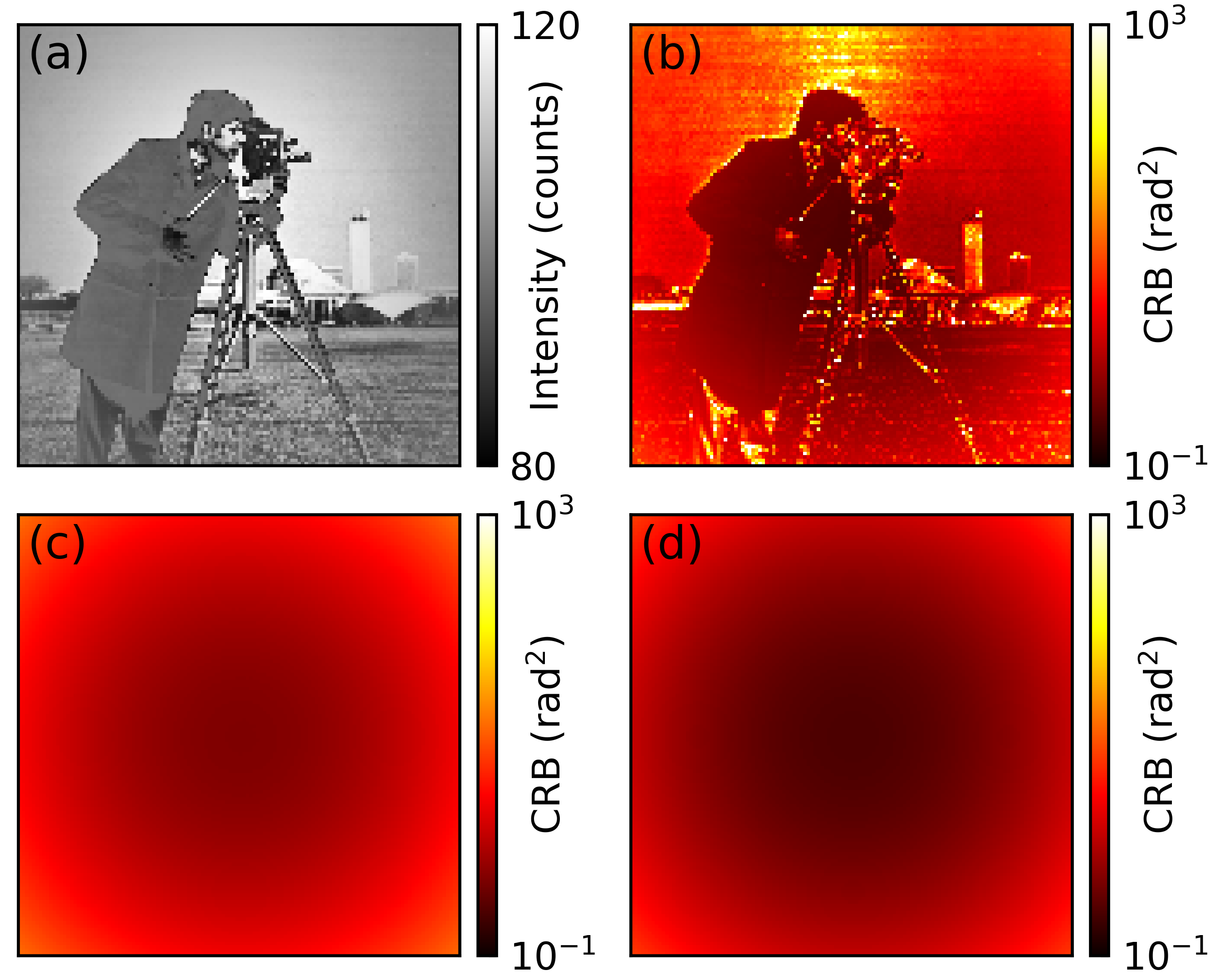}
	\caption[holo]{(a)~Simulated intensity in the camera plane for on-axis interferometric imaging with a reference beam that is phase-shifted by $\pi/2$, and (b)~resulting CRB. (c)~CRB for phase-shifting interferometric imaging using four images that are taken with a reference beam that is phase-shifted by $0,\pi/2,\pi$ and $3\pi/2$. (d)~CRB for the optimized scheme: The external reference is shaped according to the sample, such that optimal phase estimation precision is achieved across the FOV.
	}
	\label{fig:external_ref}
\end{figure}

In order to test the precision that can be achieved in such case, we generate a $128 \times 128$ complex-valued object with a phase distribution produced from the "cameraman" test image (Fig.~\ref{fig:setup}b) and with a Gaussian intensity profile (Fig.~\ref{fig:setup}c). For the calculations, we set the number of photons per frame and unit area $n_j$ to $1$ in the center of the Gaussian profile, so that one can easily deduce the CRB for other values of $n_j$ as the CRB scales with $1/n_j$. Fig.~\ref{fig:external_ref}a shows the simulated intensity distribution in the camera plane. By evaluating partial derivatives of the intensity with respect to each parameter using a finite-difference scheme, the Fisher information matrix can then be calculated according to Eq.~\eqref{general FI single parameter}, as shown in Code~1~\cite{code}. The resulting CRB (in rad$^2$) is shown in Fig.~\ref{fig:external_ref}b. In certain regions of the sample in which the phase-matching condition is fulfilled, the optimal precision is reached, resulting in a CRB equal to $1/(4n_j)=0.25$\,rad$^2$. However, the CRB strongly increases in other regions, in which the phase-matching condition is not fulfilled. We observe that this lower bound on the variance (calculated here using $n_j=1$ in the center of the FOV) can become much larger than $\pi^2$, which means that meaningful estimations of the phase cannot be achieved with $n_j\leq 1$. This has important practical implications: Even if $n_j \simeq 10^4$ (as can be achieved with typical complementary metal-oxide-semiconductor (CMOS) detectors), the CRB could be as high as $0.1$\,rad$^2$, preventing the high-speed detection of small phase shifts as induced by e.g. neuronal activity~\cite{Ling2018} or viral infection~\cite{Huang2017}. 

This problem can be addressed with phase-shifting interferometric schemes, in which the object field is successively interfered with $N\geq3$ reference fields that are phase shifted by $2u\pi/N$, where $u=0, \ldots, N-1$. As the Fisher information is additive for independent measurements, this procedure yields an averaging effect, resulting in a CRB equal to $1/(2 n_j)$, with an increase by a factor of $2$ as compared to the optimal limit (see Supplementary Section~S2). Consequently, the resulting spatial distribution of the CRB is now dictated by the Gaussian distribution of the number of incoming photons, as shown in Fig.~\ref{fig:external_ref}c. Note that the same averaging effect can also be achieved with off-axis interferometric imaging~\cite{Leith1963WavefrontObjects}. In such scheme, however, independent measurements for different values of $\alpha_k$ are obtained by oversampling the field in the camera plane instead of by taking several images. 

%In contrast, in off-axis interferometric imaging \cite{Leith1963WavefrontObjects}, the precision of both phase and absorption estimations is dictated only by the number of incident photons. In this scheme, the object field is interfered with a tilted plane wave, such that the period of the resulting interference pattern is smaller than the smallest features encoded in the object wave. Due to this oversampling, each phase value $\phi_j$ can be estimated from values measured by several camera pixels associated with different values of $\alpha_k$. This yields an averaging effect, resulting in a CRB that is increased by a factor of $2$ as compared to the optimal limit (see Supplementary Section III). The CRB relative to the precision of phase estimations is thus equal to $1/(2 n_j)$. Consequently, the resulting spatial distribution of the CRLB is now dictated by the Gaussian distribution of the number of incoming photons, as shown in Fig.~\ref{fig:external_ref}c. 

In order to realize optimal phase estimations, the phase-matching condition expressed by Eq.~\eqref{eq:phase_condition} has to be fulfilled throughout the entire FOV. This can be achieved with a SLM by spatially modulating either the reference wave (to modulate $\alpha_k$) or the object wave (to modulate $\phi_k$), yielding the optimal CRB of $1/(4 n_j)$ for phase estimations (Fig.~\ref{fig:external_ref}d). This strategy requires prior knowledge about the object and, at first glance, it may seem of little use to perform an optimal measurement on a known object. In practice, however, coarse knowledge is sufficient to initialize a measurement close to the CRB (see Supplementary Section~S3). The additional number of photons required for coarse initialization is small, and can be neglected in the analysis (especially for continuous measurements, for which initialization is only needed in the beginning of the acquisition.). For instance, such approach is routinely implemented in the context of single-mode interferometry such as in the LIGO experiment in which, apart from shot noise, various other noise sources are taken into account to find the optimal operational point~\cite{Bond2016InterferometerDetection}. Starting from an unknown object, an iterative strategy can be devised, thus progressively approaching the optimal precision limit~\cite{Higgins2007a}.%As we will see later, LowPhi realizes such a measurement using an internal reference wave \cite{Juffmann2018}.

\begin{figure}
\centering
\includegraphics[width=\columnwidth]{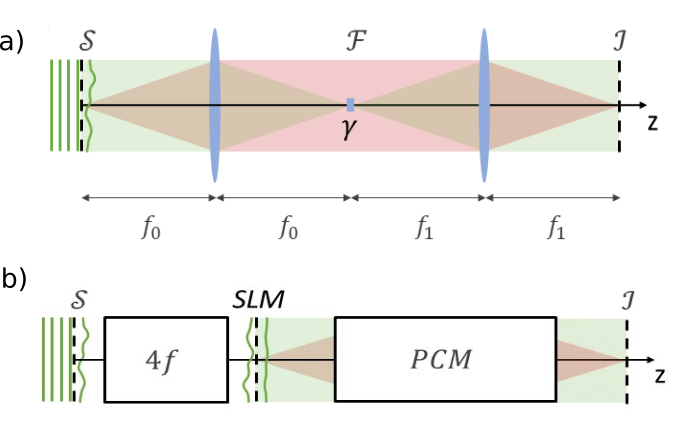}
\caption{(a)~Schematics of a phase contrast microscope (PCM). The sample $\mathcal{S}$ is imaged onto the image plane $\mathcal{I}$ using a modified 4f setup, in which the unscattered light (green) is phase-shifted by an angle $\gamma$ with respect to the scattered light (red). For an ordinary PCM, the phase shift is realized in the Fourier plane $\mathcal{F}$ and its value is set to $\gamma = \pm \pi/2$. (b)~Schematics of a LowPhi setup. The sample $\mathcal{S}$ is imaged onto an SLM using a 4f setup (note that the order could be reversed). The SLM is used to subtract an estimated phase distribution from the object wave, which is then imaged using the PCM scheme shown above.}
\label{fig:PCM}
\end{figure}

\section{Phase microscopes with an internal reference}\label{sec:internalref}

\begin{figure}
	\centering
	\includegraphics[width=\columnwidth]{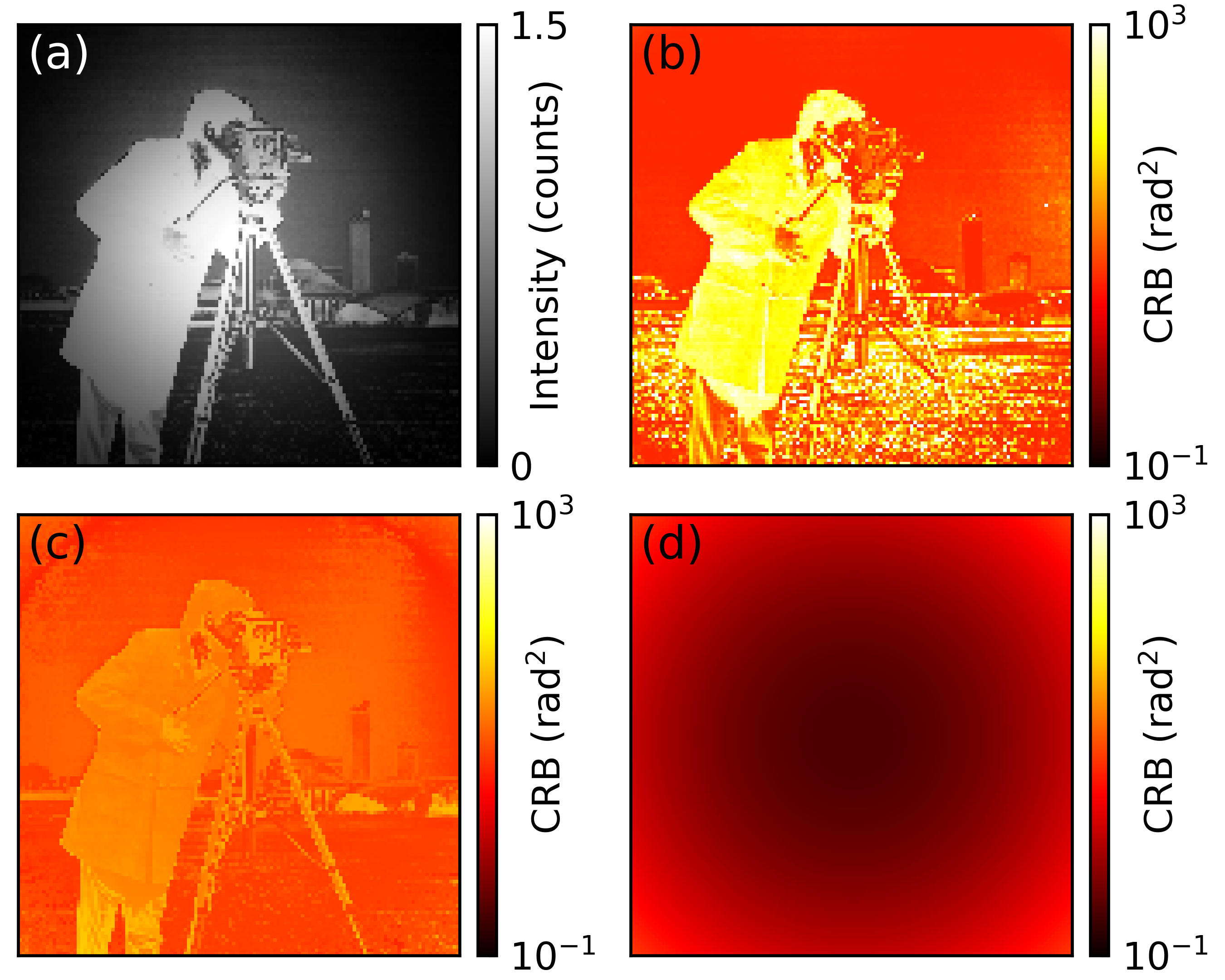}
	\caption[holo]{(a)~Simulated intensity in the camera plane for PCM and (b)~resulting CRB. (c)~CRB for SLIM: Four images are taken with an internal reference that is phase-shifted by $0,\pi/2,\pi$ and $3\pi/2$. (d)~CRB for Lowphi: An SLM is used to subtract an estimated phase distribution from the object wave, such that optimal phase estimation precision is achieved across the FOV.
	}
	\label{fig:internal_ref}
\end{figure}

Techniques based on an external reference are experimentally often limited by vibrations, which lead to an unstable phase difference between reference and object wave. This constraint led to the widespread application of phase microscopy using an internal reference wave in a stable common-path geometry. Already in the 30s, Zernike realized that a wavefront transmitted by a weakly-contrasted object can be understood as the superposition of an unscattered plane wave and a scattered wave, which led him to invent phase contrast microscopy (PCM)~\cite{Zernike1942}. In an ordinary PCM, a typically ring- or disc-shaped phase mask is placed in the Fourier plane of the 4f imaging system to shift the phase between the scattered and unscattered wave by $\pm \pi/2$. However, while this strategy yields the optimal precision for phase estimations in the case of weakly-contrasted objects, this is not the case for strongly-contrasted objects. We can numerically model a PCM by applying a fast Fourier transform (FFT) algorithm to the phase object represented in Fig.~\ref{fig:setup}b and by shifting the phase of the central pixel by $\gamma=\pi/2$. The resulting intensity distribution in the camera plane (Fig.~\ref{fig:internal_ref}a) and the associated sub-optimal CRB (Fig.~\ref{fig:internal_ref}b) remarkably differ from the results shown in Fig.~\ref{fig:external_ref}a and Fig.~\ref{fig:external_ref}b, where we considered an external reference wave shifted by $\pi/2$. Indeed, highly scattering regions deplete the internal reference (i.e. the light passing through the phase mask), making phase measurements highly inefficient. Another important consideration is that, with an internal reference, measurements are insensitive to the global phase of the object wave. Thus, the Fisher information matrix is singular unless this parameter is excluded from its definition. Such consideration is especially important when considering a situation in which spatial frequencies beyond the numerical aperture of the optical system are lost, and in which only absorption information can be accessed for low spatial frequencies that fall within the phase disc (see Supplementary Section~S4).

It is instructive to compare these results with imaging schemes that rely on phase stepping. One example is spatial light interference microscopy (SLIM)~\cite{Wang2011SpatialSLIM}, a scheme that is similar to standard PCM but which includes the option to control the phase shift $\gamma$. Several measurements are performed by successively changing the value of $\gamma$ to $ 0$, $\pi/2$, $\pi$ and $3\pi/2$, as implemented in Ref. \cite{Wang2011SpatialSLIM}. While we showed that the CRB depends only on the number of incident photons for a phase-shifting interferometric scheme with an external reference (see Fig. \ref{fig:external_ref}c), this is not the case for an internal reference wave. This can be seen in Fig.~\ref{fig:internal_ref}c, which shows strong variations in the CRB across the FOV. This effect is again due a depletion of the internal reference for strongly scattering regions, as already discussed in our analysis of PCM. %Note that this problem can be alleviated by increasing the size of the phase disc, which happens at the expense of losing all knowledge about phase changes happening at spatial frequency components that fall within the phase disc, and of introducing well-known image artefacts like halos \tjcomment{reference to suppl}. 

This problem can be addressed using a recent method called local optimization of wavefronts for phase imaging (LowPhi)~\cite{Juffmann2018}. In this scheme, the object is imaged onto a SLM and the resulting wavefront is then imaged using traditional PCM (Fig. \ref{fig:PCM}b). The system is first initialized, such that the SLM approximately cancels phase variations induced by the object. If the error in the initialization is small, then the following PCM is an optimal measurement scheme for these deviations. Fig.~\ref{fig:internal_ref}d shows the resulting CRB, which is indeed optimal over the entire FOV. This demonstrates that one does not need an external reference beam (as considered in Fig.~\ref{fig:external_ref}d) to reach the optimal precision for phase estimations. In both cases, however, optimal phase estimations come at the price of the required initialization (the photon budget of which can be neglected for continuous measurements) and technical issues regarding SLM phase noise~\cite{Moser2019Model-basedModulators}. Moreover, measurements are then fully insensitive to the amplitude of the object wave, and one would have to replace the PCM after the SLM by a standard bright-field microscope in order to perform optimal absorption estimations. 

\section{Conclusion}\label{sec:conclusion}

To summarize, we showed that the Cram\'{e}r-Rao bound is a powerful tool to investigate the estimation precision of phase microscopes. The general framework that we introduced here enables the analysis of any classical phase microscope that can be described by a linear transformation. This allows to assess whether a chosen technique can in principle provide the precision required for challenging tasks, such as the high-speed detection of small phase shifts induced by neuronal activity~\cite{Ling2018} or viral infection~\cite{Huang2017}. It can also allow one to verify whether the precision of phase estimates is limited by shot noise, or whether this precision is degraded by additional noise sources (e.g. due to mechanical instabilities) which could easily be described using the same formalism. 

Importantly, our analysis yields the necessary conditions for reaching the optimal precision in phase estimations (as well as those needed to perform optimal absorption estimations). A critical requirement for reaching the optimal phase precision is a phase-matching condition between object wave and reference wave, which can either be internal or external. With this theoretical framework, we analyzed the precision achievable with different experimental configurations, evidencing that wavefront shaping can enable optimal phase (or absorption) estimation precision, even in the absence of an external reference wave (e.g. in LowPhi~\cite{Juffmann2018}). In practice, such approaches based on wavefront shaping are currently limited by SLM phase noise, therefore requiring alternative approaches to phase microscope optimization based on e.g. generalized phase contrast~\cite{Gluckstad2009} or deep learning~\cite{8667888}. Interestingly, these alternatives approaches could also be benchmarked using the Cram\'{e}r-Rao analysis presented here.

Further work will be needed to assess the performance of minimum variance unbiased estimators relative to these bounds, and to perform experimental measurements with a precision reaching the theoretical limit. Finally, note that the formalism can be extended to schemes that further improve the precision of phase microscopy using cavities~\cite{Nimmrichter2018, Mader2015a}, and can also be applied to electron phase microscopy techniques~\cite{Danev2017, Ophus2016EfficientInterferometry}.

%Fixing the phase of the reference beam thus leads to an estimation precision that varies across the field of view and that can even diverge in certain regions of the object. In schemes based on an external reference, continuous or discrete phase stepping schemes can alleviate this problem, yielding an averaged estimation precision for both phase and absorption. For schemes based on an internal reference wave, this is only true as long as the presence of the sample does not deplete the intensity in the internal reference. Our analysis further shows that wave-front shaping (e.g. Lowphi \cite{Juffmann2018}) can enable optimal phase (or absorption) estimation precision in both internal and external phase reference schemes. 

%Deep learning regularization: we lose independence between the parameters to estimate, the reconstruction becomes dataset-dependent

\medskip

\noindent{\textbf{Funding.}} %This work was supported by the 
DB acknowledges support from the Netherlands Organization for Scientific Research NWO (Vici 68047618 and Perspective P16-08). TJ and DM acknowledge support from the ERC Micromoupe Grant 758752.

\medskip

\noindent{\textbf{Acknowledgments.}} The authors thank Allard P.\ Mosk for insightful discussions.

\medskip

\noindent{\textbf{Disclosures.}} The authors declare no conflicts of interest.

\medskip

\noindent See Supplemental Material for supporting content. The Python code for this project can be found at: https://github.com/JuffmannLab/PhaseMicroscopyCRB

% Bibliography
\bibliography{refs}

% Full bibliography added automatically for Optics Letters submissions; the following line will simply be ignored if submitting to other journals.
% Note that this extra page will not count against page length
% \bibliographyfullrefs{refs}

%Manual citation list
%\begin{thebibliography}{1}
%\bibitem{Zhang:14}
%Y.~Zhang, S.~Qiao, L.~Sun, Q.~W. Shi, W.~Huang, %L.~Li, and Z.~Yang,
%  \enquote{Photoinduced active terahertz metamaterials with nanostructured
%  vanadium dioxide film deposited by sol-gel method,} Opt. Express \textbf{22},
%  11070--11078 (2014).
%\end{thebibliography}

\newpage
\appendix

\section{Derivation of the optimal Cram\'{e}r-Rao bound for absorption estimations}

Instead of estimating the set of parameters $\phi=(\phi_1,\dots,\phi_p)$, one may want to estimate the set of parameters $n=(n_1,\dots,n_p)$ in order to characterize an absorptive sample. In the same way that we obtained the Fisher information matrix $\mathcal{J}(\phi)$ for phase estimations, we now consider the Fisher information matrix $\mathcal{J}(n)$ for absorption measurements. In the shot-noise limit and assuming that the values measured by all camera pixels are statistically independent, we obtain
\begin{equation}
 [\mathcal{J}(n)]_{ij} 
 = \Delta t \sum_{k=1}^d
 \frac{1}{I_k^{\rm{det}}}
 \left( \frac{\partial I_k^{\rm{det}}}{\partial n_i} \right)
 \left( \frac{\partial I_k^{\rm{det}}}{\partial n_j} \right) \; .
\end{equation}
In order to express the derivative of the intensity $I_k^\mathrm{det}$ with respect to $\phi_j$, we can use the following chain rule:
\begin{equation}
 \label{chain rule}
 \frac{\partial I_k^{\rm{det}}}{\partial n_j} = \left(\frac{\partial I_k^{\rm{det}}}{\partial \mathcal{A}_j}\right) \left( \frac{\partial \mathcal{A}_j}{\partial n_j} \right) \; .
\end{equation}
This yields
\begin{align}
 \frac{\partial I_k^{\rm{det}}}{\partial n_j} 
 &= \frac{1}{2 \sqrt{\Delta t n_j}} \left[ 
 (E_k^{\rm{det}})^* h_{kj} e^{i\phi_j}
 + E_k^{\rm{det}} h_{kj}^* e^{-i\phi_j} \right] \nonumber\\
 &= \frac{1}{\sqrt{\Delta t n_j}} \text{Re}\left[
 (E_k^{\rm{det}})^* h_{kj} e^{i\phi_j} \right] \; .
\end{align}
The elements of the Fisher information matrix can thus be explicitly expressed as follows:
\begin{equation}
\begin{split}
 \label{explicit fisher information matrix n}
 [\mathcal{J}(n)]_{ij}
 = &\frac{1}{n_j} \sum_{k=1}^d \frac{ 1}{|E^{\rm{det}}_k|^2} \\ & \times \text{Re}\left[
 (E_k^{\rm{det}})^* h_{ki} e^{i\phi_i} \right]
 \text{Re}\left[
 (E_k^{\rm{det}})^* h_{kj} e^{i\phi_j} \right] \;  .
 \end{split}
\end{equation}
Writing $E^{\rm{det}}_k = |E^{\rm{det}}_k| e^{i \alpha_k} $, the diagonal elements of the Fisher information matrix are expressed by:
\begin{equation}
 [\mathcal{J}(n)]_{jj}= \frac{1}{n_j} \sum_{k=1}^d |h_{kj}|^2 \cos^2(\phi_j+\beta_{kj}-\alpha_k) \; .
 \label{diag_elements n}
\end{equation}
Considering that $ \cos^2(\phi_j + \beta_{kj} - \alpha_k) \leq 1$ and $\sum_k |h_{kj}|^2 \leq 1$, we have the following inequality:
\begin{equation}
 [\mathcal{J}(\phi)]_{jj}\leq \frac{1}{n_j} \; ,
 \label{diag_elements_bound_n}
\end{equation}
which is saturated if $H$ is unitary and if the following phase-matching condition is satisfied:
\begin{equation}
 \phi_j + \beta_{kj} - \alpha_k = m \pi \; ,
 \label{phase_matching_amplitude}
\end{equation}
where $m$ can be any integer. This must hold for all $k, j$ for which $n_j h_{kj} \neq 0$. By comparing Eq.~\eqref{phase_matching_amplitude} to the corresponding one for phase estimation in the main text, it is clear that the condition required to perform optimal phase estimations is complementary to the condition required to perform optimal absorption estimations.

To summarize, we can write the following chain of inequalities for absorption estimations:
\begin{equation}
 \operatorname{Var}(\hat{n}_j) \geq
 [ \mathcal{J}^{-1}(n) ]_{jj} \geq
 [\mathcal{J}(n)]_{jj}^{-1} \geq
 n_j \; ,
\end{equation}
where the optimal bound simply corresponds to the variance of shot-noise limited measurements in a direct imaging configuration of an absorptive sample with known phase contrast. 

Note that we have treated phase and absorption estimations separately. In general, the number of unknowns cannot exceed the number of measurements for the Fisher information matrix to be well-conditioned. Precisely determining both phase and absorption entails estimating $2p$ parameters, and the Fisher information matrix can be well-conditioned only if the number $d$ of measurements satisfies $d \geq 2p$. This can be achieved either in a single shot (i.e. in off-axis interferometry) or with multiple acquisitions (i.e. in phase shifting interferometry), as will be discussed in the following section. %Using such procedure, the fundamental bounds on the variance of amplitude and phase estimations are both increased by a factor of $2$. This corresponds to a usual cost for simultaneously estimating both amplitude and phase: twice the number of photons is needed to achieve the same sensitivity as for amplitude-only or phase-only estimations. 

\section{Cram\'{e}r-Rao bound for phase-shifting and off-axis interferometry}

We now study the case of a simple phase-shifting interferometric scheme in which the object field is successively interfered with $N\geq 3$ external reference plane waves that are phase-shifted by $2u\pi/N$, where $u=0, \ldots, N-1$. We assume an ideal detection scheme so that $H$ is the identity matrix of size $p$, resulting in a diagonal Fisher information matrix. We also suppose that the intensity of the reference wave is high everywhere in the field of view, so that $\alpha_k \simeq \arg (E^{\rm{ref}}_k)$. Then, considering that the $N$ successive measurements are statistically independent, we can add the Fisher information associated with each measurement. Taking an integration time of $\Delta t/N$ for each measurement to keep the total photon number incident on the sample constant, we can use Eq.~(8) of the main text to calculate the diagonal elements of the Fisher information matrix, which reads
\begin{equation}
 [\mathcal{J}(\phi)]_{jj}=  \frac{4n_j}{N} \sum_{u=0}^{N-1} \sin^2(\phi_j + 2u\pi/N) \; . 
 \label{diag_elements_phase_shifting}
\end{equation}
Making use of trigonometric identities, we readily obtain
%\begin{equation}
%    [\mathcal{J}(\phi)]_{jj} = \frac{2n_j}{N} \sum_{u=0}^{N-1} \left[ 1 - \cos(2\phi_j + 4u\pi/N) \right] \; .
%\end{equation}
%This expression can be expanded into
\begin{equation}
    [\mathcal{J}(\phi)]_{jj} =2 n_j - \frac{n_j}{N} \sum_{u=0}^{N-1} \left[ e^{2i (\phi_j + 2u\pi/N)} + \mathrm{c.c.} \right] \; ,
\end{equation}
where $\mathrm{c.c.}$ denotes the complex conjugate of the preceding term. Computing the geometric series, this yields
\begin{equation}
 [\mathcal{J}(\phi)]_{jj}= 2 n_j - \frac{n_j}{N} \left[e^{2 i \phi_j} \left( \frac{1-e^{i 4 \pi}}{1-e^{i 4 \pi/N}}\right) + \mathrm{c.c.} \right] \; .
\end{equation}
Since $N \geq 3$, we finally obtain
\begin{equation}
 [\mathcal{J}(\phi)]_{jj}= 2 n_j  \; .
     \label{diag_elements_phase_shifting_2}
\end{equation}
The resulting Cram\'{e}r-Rao bound is $1/(2 n_j)$, which is two times worse than the optimal Cram\'{e}r-Rao bound for phase estimations. Note, however, that the optimal scheme for phase estimations yields no absorption information. In contrast, the diagonal elements of the Fisher information matrix relative to absorption estimations is non-zero for the phase-shifting interferometric scheme. Starting from Eq.~\eqref{diag_elements n}, it can be easily shown that
\begin{equation}
 [\mathcal{J}(n)]_{jj}= \frac{1}{2n_j} \; .
 \label{diag_elements n_phase_shifting}
\end{equation}
The resulting Cram\'{e}r-Rao bound is then $2 n_j$, which is two times larger than the optimal Cram\'{e}r-Rao bound for absorption estimations. 

Similar results can also be obtained for an off-axis interferometric imaging scheme. In such scheme, the object field is interfered with a tilted plane wave, such that the period of the resulting interference pattern is smaller than the smallest features encoded in the object wave. Due to this oversampling, each phase value $\phi_j$ can be estimated from values measured by $q$ camera pixels associated with different values of $\alpha_k$ ranging from $0$ to $2 \pi$. This yields an averaging effect similar to what was exposed earlier in the case of the phase-shifting interferometric scheme, resulting in a Cram\'{e}r-Rao bound of $1/(2 n_j)$ for the estimation of $\phi_j$ (phase estimations) and of $2 n_j$ for the estimation of $n_j$ (absorption estimations).

\section{Optimal phase microscopes with incomplete knowledge of phase objects}

\begin{figure}[b!]
	\centering
	\includegraphics[width=7.5cm]{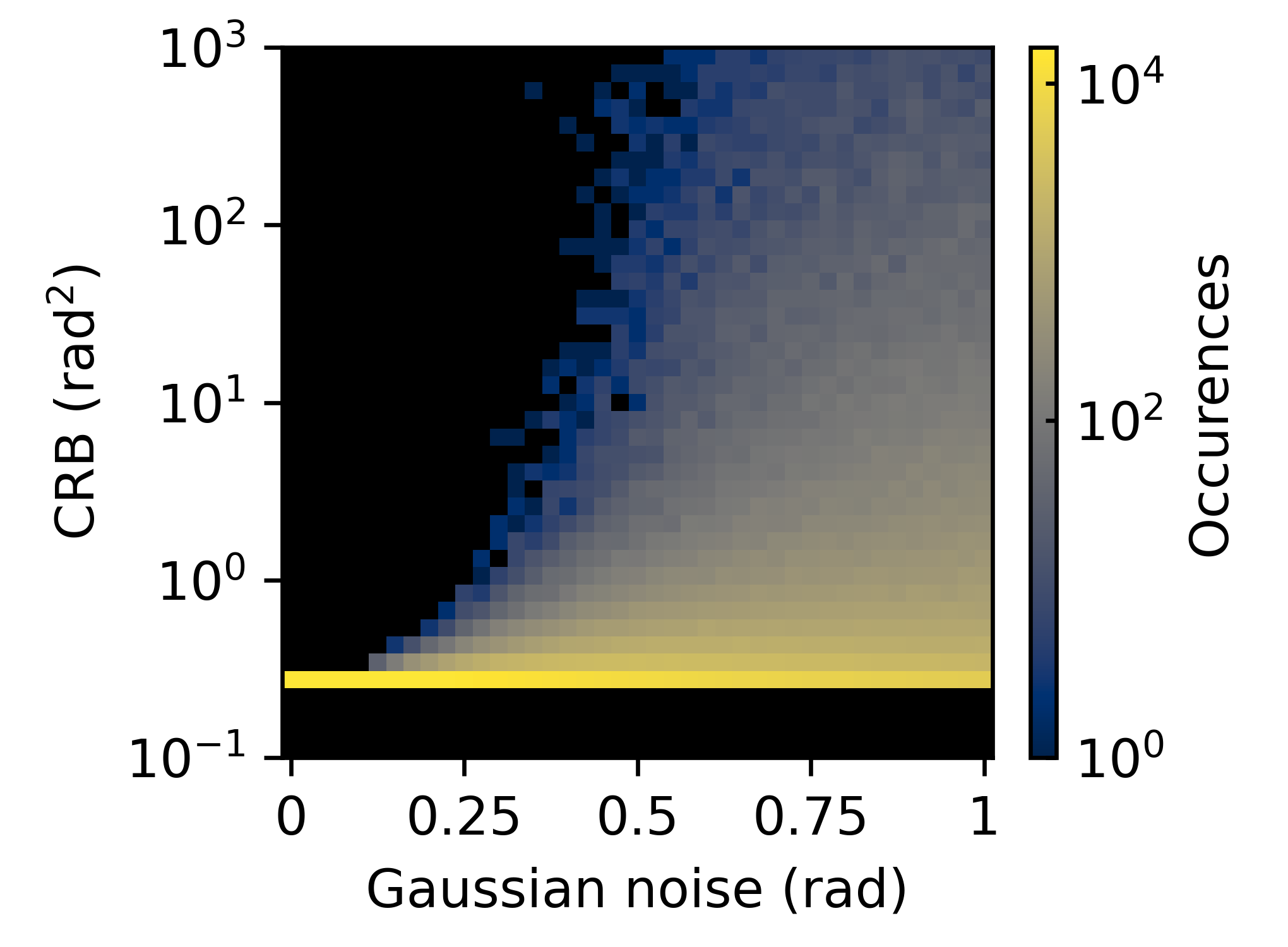}
	\caption[holo]{Stack of histograms of CRB as a function of the standard deviation of a Gaussian uncertainty regarding the true phase distribution in the object plane. These calculations were performed for a $128\times128$ phase object and assuming an ideal detection setup with a high-intensity optimally-shaped reference beam. No occurrences were observed in the black areas of the figure. }
	\label{fig:initialization}
\end{figure}

Optimal detection schemes (with or without external reference) require prior knowledge of the entire sample in order to effectively turn a strongly-contrasted object into a weakly-contrasted object. If the prior knowledge is incomplete, the phase-matching condition will then only be approximately satisfied. To determine the influence of such incomplete knowledge, we consider an optimal detection scheme with a uniform intensity distribution ($n_j=1$ in the whole field of fiew), with a high-intensity reference beam ($|E^\mathrm{ref}_k|^2/|E^\mathrm{obj}_k|^2=100$ in the entire field of view) and with a phase such that $\arg (E^\mathrm{ref}_k)=\pi/2+\phi_k+W_k$, where $W_k$ follows a centered Gaussian distribution of variance $\sigma_\mathrm{g}^2$. We then vary the standard deviation of the noise from $0$ to $1$\,rad and, for each value of $\sigma_\mathrm{g}$, we plot the histogram of the CRB obtained within the field of view. As can be seen in Fig.~\ref{fig:initialization}, close to optimal estimations can be performed for $\sigma_\mathrm{g}<\pi/8$, with only a few occurrences characterized by a sub-optimal CRB. Note that it is this regime in which PCM works efficiently; spatial light interference microscopy (SLIM) is then a potential option as an initialization before using an optimal scheme such as Lowphi to perform dynamic precise phase estimations.

\section{Information loss and singular Fisher information matrices}

We now discuss important effects that arise when we consider that the finite numerical aperture (NA) of the imaging system blocks all high spatial frequencies in the image plane, and that the finite size of the phase mask prevents phase estimations for low spatial frequencies (only absorption estimations could be performed for such frequencies). Taking into account these two effects, the Fisher information matrix of phase microscopes is necessarily singular, and the associated Cram\'{e}r-Rao bound therefore cannot be calculated without further analysis. In order to understand why a direct inversion of the Fisher information matrix is not possible, it is relevant to use the Fourier basis for representing the parameters $\phi$. Thus, we consider the estimation of a new set of parameters $\xi=W \phi $, where $W$ is the (unitary) discrete Fourier transform (DFT) matrix which is used to approximate the Fourier transform operator. The Fisher information matrix for the parameters $\xi$, calculated as $\mathcal{J}(\xi)=W \mathcal{J}(\phi) W^\dagger$, has zeros for both low and high spatial frequencies, resulting in a singular Fisher information matrix.

\begin{figure}[b!]
	\centering
	\includegraphics[width=\columnwidth]{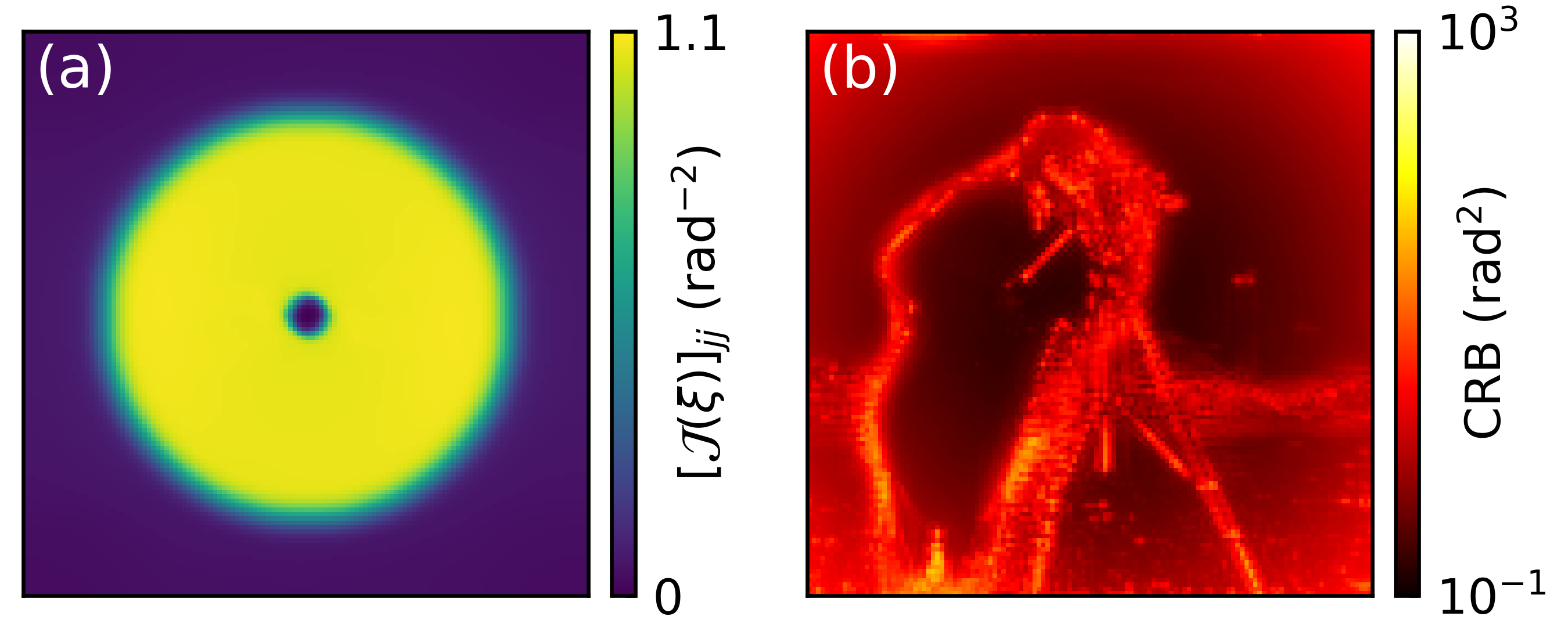}
	\caption{(a)~Diagonal elements of the Fisher information matrix $\mathcal{J}(\xi)$, expressed in the Fourier basis. (b)~Resulting Cram\'{e}r-Rao bound obtained by truncating $\mathcal{J}(\xi)$ for low and high spatial frequencies.}
	\label{fig:truncatedFI}
\end{figure}

As an example, we study the PCM configuration described in the manuscript but with a finite NA and a finite phase mask. To this end, we suppose that the area of a pixel in the detection plane is $\lambda_0^2/4$ where $\lambda_0$ is the wavelength of the incident light, so that the field is sampled at the Nyquist frequency. We further assume to use an Olympus microscope objective with a numerical aperture $\mathrm{NA}=0.75$ and a $\times 40$ magnification, along with a phase disc whose radius corresponds to the width of a commonly-used Ph2 phase ring. The finite numerical aperture of the microscope objective effectively blocks all spatial frequencies $k/k_0>0.75$ where $k$ is the magnitude of the transverse component of the wavevector and $k_0=2 \pi/\lambda_0$. Furthermore, the effect of the phase ring is to shift all spatial frequencies $k/k_0<0.07$ by $\gamma=\pi/2$. The resulting distribution for the diagonal elements of $\mathcal{J}(\xi)$ is shown in Fig.~\ref{fig:truncatedFI}a. Note that a zero in a diagonal element of $\mathcal{J}(\xi)$ implies that a whole row and a whole column of $\mathcal{J}(\xi)$ is zero. This demonstrates that the Fisher information is suppressed for high spatial frequencies (due to the finite NA of the detection system) as well as for low spatial frequencies (due to the finite size of the phase mask). In the limit of an idealized setting with infinite NA and with a phase mask assumed to be infinitely small, this has the well-known consequence that only a uniform global phase factor cannot be determined from the intensity distribution in the camera plane, as discussed in the manuscript. 

However, in order to continue the analysis, we can assume that the value of $\xi_j$ associated with the missing spatial frequencies are provided as \textit{a priori} information, or that it is known that the object does not feature such frequencies. In this case, we can reduce the dimensions of $\mathcal{J}(\xi)$ and $W$ by removing the lines and columns associated with these parameters. This procedure results in the construction of the truncated matrices $\tilde{\mathcal{J}}(\xi)$ and $\tilde{W}$, which satisfy $\tilde{\mathcal{J}}(\xi)=\tilde{W} \mathcal{J}(\phi) \tilde{W}^\dagger $. The resulting Fisher information matrix $\tilde{\mathcal{J}}(\xi)$ is now invertible, which allows us to write the following Cram\'{e}r-Rao inequality
\begin{equation}
\operatorname{Var}(\hat{\phi}_j) \geq \left[\tilde{W}^\dagger \tilde{\mathcal{J}}^{-1}(\xi) \tilde{W}\right]_{jj} \; .
\end{equation}
Note that, due to the addition of \textit{a priori} information, the Cram\'{e}r-Rao bound can be smaller than the fundamental limit expressed by $1/(4 n_j)$ in certain regions of the sample. 

In the numerical results shown in Fig.~4b of the manuscript, the NA was assumed to be infinite, and the phase disc was assumed to be one pixel in size, so the Fisher information matrix had to be truncated by one column and one line before inversion. It is interesting to compare these results to the more realistic case of a finite size phase-disc and limited NA. Fig.~\ref{fig:truncatedFI}b shows the Cram\'{e}r-Rao bound obtained by considering the Fisher information matrix whose diagonal elements are shown in Fig.~\ref{fig:truncatedFI}a, and by truncating it to remove all zero spatial frequencies. Remarkably, the larger phase mask and the underlying assumption of \textit{a priori} knowledge about these frequencies lead to a reduced average Cram\'{e}r-Rao bound. Moreover, we can recognize a Fourier-filtered version of the cameraman on the map in Fig. \ref{fig:truncatedFI}b, which confirms that low and high spatial frequencies are suppressed from the measurements. %Finally, we can see that the Cram\'{e}r-Rao bound increases in regions where the sample-induced phase-shifts change with high spatial frequency, which is due to artefacts (e.g. halos) often present in phase microscopy. 

\end{document}